# A New Catalog of Head–Tail Radio Galaxies from the VLA FIRST Survey

Tapan K. Sasmal[1], Soumen Bera[1], Sabyasachi Pal[2,3], and Soumen Mondal[1]
[1] Department of Physics, Jadavpur University, Kolkata, 700032, India; tapanksasmal.phys.rs@jadavpuruniversity.in
[2] Midnapore City College, Kuturia, Bhadutala, Paschim Medinipur, West Bengal, 721129, India
[3] Indian Centre for Space Physics, 43 Chalantika, Garia Station Road, 700084, India

## Abstract

The head–tail (HT) morphology of radio galaxies is seen for a class of radio sources where the primary lobes are being bent in the intercluster weather due to strong interactions between the radio jets and their respective intracluster medium. A systematic search has been carried out for new HT radio galaxies from the Very Large Array Faint Images of the Radio Sky at Twenty-Centimeters survey database at 1400 MHz. Here, we present a catalog of 717 new HT sources, among which 287 are narrow-angle tail (NAT) sources whose opening angle between the two lobes is less than 90°, and 430 are wide-angle tail (WAT) whose the opening angle between the two lobes is greater than 90°. NAT radio sources are characterized by tails bent in a narrow "V"-like shape; the jet bending in the case of WAT radio galaxies are such that the WATs exhibit wide "C"-like morphologies. Optical counterparts are found for 359 HT sources. We report HT sources with luminosity ranges $10^{38} \leqslant L_{1.4\,\mathrm{GHz}} \leqslant 10^{45}$ erg s$^{-1}$ and redshifts up to 2.01. The various physical properties of these HT sources are mentioned here. Some statistical studies have been done for this large number of newly discovered HT sources.

*Unified Astronomy Thesaurus concepts:* Active galactic nuclei (16); Catalogs (205); Interferometric correlation (807); Quasars (1319); Radio continuum emission (1340); Surveys (1671); Galaxy clusters (584); Tailed radio galaxies (1682)

*Supporting material:* machine-readable tables

## 1. Introduction

The morphologies of radio galaxies show a number of finite, known structures of classes and subclasses along with the typical one. A typical radio galaxy has two oppositely directed radio jets from a central core. These radio-jet structures and their relative orientations give rise to a set of new radio morphology classes. A head–tail (HT) radio galaxy (Ryle & Windram 1968; Rudnick & Owen 1976; Blanton et al. 2000; Proctor 2011; Dehghan et al. 2014; Missaglia et al. 2019; Patra et al. 2019) is one of such classes, where the two radio jets are seen to be bent in the same direction on the radio map and give the overall shape of the radio source a "C", "L", or "V" shape. Ryle & Windram (1968) first described HT radio sources. The HT sources are also known as bent-tailed (BT) radio sources, as the tail is supposed to be bent. The bending of the two jets forms a certain angle (<180°) at the center of the source. Owen & Rudnick (1976) first classified this type of galaxy into two categories depending upon their degree of bending and luminosity, wide-angle tail (WAT) and narrow-angle tail (NAT). The technical definitions of the classes are as follows: WATs are identified as bent radio galaxies having intermediate radio luminosity $10^{42} \leqslant L_{\mathrm{WAT}} \leqslant 10^{43}$ erg s$^{-1}$ for a radio window 10$^7$ to 10$^{11}$ Hz with central dominant galaxies in rich galaxy clusters (O'Donoghue et al. 1990, 1993). On the other hand, when the two tails are asymptotic to roughly parallel trajectories away from the head, they are known as NAT radio galaxies (Rudnick & Owen 1976). NAT sources are commonly found in cluster environments. HT sources with an opening angle between the radio jets greater or equal to 90° are known as WAT, while sources with an angle less than 90° are termed as NAT. As a criterion to define WATs and NATs, we have used the angle between the jets. The radio source 3C 465 is one of the best-known examples of a WAT-class radio source (Eilik et al. 1984; Eilek & Owen 2002; Hardcastle et al. 2005). On the other hand, the radio source NGC 1265 is used as a good example of a NAT source (Ryle & Windram 1968; O'Dea & Owen 1986).

These types of objects are usually found in the environment of rich clusters of galaxies (Burns 1990). They are supposed to be moving through the intracluster medium (ICM) with sufficient velocities for the tails to bend by the action of ram pressure (O'Dea 1985). In this way, the primary origin for the bending of the jet is supposed to be the dynamic pressure that pushes back the jets; and the reason for this pressure is the high-velocity motion of the associated galaxy through the surrounding ICM. This "ram-pressure" model was first mentioned by Begelman et al. (1979). Later this ram-pressure model was well explained by many authors (Baan & McKee 1985; Vallee 1981). A buoyancy force was also invoked to understand the bending of the jets. When the material density of the radio jets is less than the density of the surrounding medium, the buoyancy force comes into action; it pushes the lobes to the regions of the ICM where the density of the jet is equal to that of the surrounding medium and thus bends the jets (Gull & Northover 1973; Sakelliou et al. 1996). There is an alternative theory wherein the distortion of jets occurs as the result of the high velocity of the ICM due to a merger of clusters (Burns 1990; Roettiger et al. 1996; Blanton et al. 2000; Mao et al. 2009; Dehghan et al. 2014). These models can be used to probe unknown galaxy clusters with the presence of a HT source.

Based on morphology and radio power, Fanaroff–Riley classified radio sources mainly in two categories: Fanaroff–Riley type I (FR I), which have low power (less than

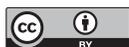

$5 \times 10^{25}$ W Hz$^{-1}$ at 1440 MHz) with bright cores and their lobes fading toward the edges; and Fanaroff–Riley type II (FR II), which have a power greater than $5 \times 10^{25}$ W Hz$^{-1}$ at 1440 MHz and dim cores, with edge-brightened lobes (Fanaroff & Riley 1974).

Considering the previous detection of bent radio sources, Owen & Rudnick (1976) first presented the observational result of six WATs in rich clusters of galaxies using the NRAO interferometer at 2695 MHz. A total of 11 WAT sources were presented by O'Donoghue et al. (1990) from Very Large Array (VLA) observation at 20 cm. Later, O'Donoghue et al. (1993) presented images of 11 WATs at both 6 cm and 20 cm to study the dynamics of such kinds of sources. From VLA Faint Images of the Radio Sky at Twenty-Centimeters (FIRST) survey data Blanton (2000) reported a total of 384 bent-double radio sources. Wing & Blanton (2011) studied the optical environments around double-lobed radio sources from the FIRST survey. Their samples contain 384 visually selected bent samples (from Blanton 2000), 1546 automatically selected bent samples (from Proctor 2006), 3232 straight samples (from Proctor 2006) and 3348 single-component samples from the FIRST catalog. They found that 78% of the time the visually selected bent sources were associated with clusters or rich groups. The association rate drops to 59%, 43%, and 29% for auto-bent, straight, and single-component samples, respectively. By using the 1.4 GHz Australia Telescope Large Area Survey (ATLAS), Dehghan et al. (2014) presented a catalog of 56 BT objects. In The Spitzer catalog, Paterno-Mahler et al. (2017) found 190 galaxy-cluster candidates from the study of the surrounding fields of 646 bent, double-lobed radio sources. A total of 47 WAT candidates were cataloged by Missaglia et al. (2019), using a combination of National Radio Astronomy Observatory/Very Large Array Sky Survey (NVSS), FIRST and the Sloan Digital Sky Survey (SDSS) databases. There have also been some individual studies about special types of HT radio galaxies, as found in the literature. Recently, a multifrequency property of an interacting NAT radio galaxy, J0037+18, was presented by Patra et al. (2019) using the Giant Meterwave Radio Telescope (GMRT) and VLA data. A follow-up work on the longest-known HT radio galaxy, IC 711, was undertaken by Srivastava & Singal (2020). Recently, Sasmal et al. (2022) cataloged 82 new HT sources from the LOFAR Two-meter Sky Survey first data release (LoTSS DR1) at 144 MHz frequency.

Here, we present the outcomes of a symmetric visual search for the identification of WAT and NAT sources from the FIRST data catalog at 1.4 GHz. The present paper is arranged in the following order. In Section 2 we detail the methodology of our work, which includes the search strategy for HT sources from FIRST data, finding the respective optical counterpart, etc. The results of our search are given in Section 3. In this results section the physical parameters of the newly identified HT sources, like spectral index, luminosity, FR type, etc., are estimated. The discussion and conclusion are presented in Sections 4 and 5, respectively.

In this paper the following cosmology parameters have been used: $H_0 = 67.4$ km s$^{-1}$ Mpc$^{-1}$, $\Omega_m = 0.315$, and $\Omega_{\rm vac} = 0.685$ (Aghanim et al. 2020). In the calculations, the spectral index ($\alpha$) is defined as $S_\nu \propto \nu^\alpha$, where $S_\nu$ is the flux density.

## 2. Methodology

### 2.1. The FIRST Survey Data

To deal with sources with diffuse emission, one should first use survey data with better resolution. In this scenario, we use the FIRST survey data, which utilizes the NRAO VLA in its B-configuration (VLA B; Becker et al. 1995). The FIRST survey covers a radio sky of 10,575 square degrees, which is equivalent to 25% of the entire sky. Individually, this survey includes an area of R.A. 7.0$^h$ to 17.5$^h$ and decl. −8°.0 to +57°.6, totaling 8444 square degrees in the north Galactic cap and R.A. 20.4$^h$ to 4.0$^h$, decl. −11°.5 to +15°.4, totalling 2131 square degrees in the south sky. The total FIRST sky is mapped with 3 minute snapshots covering a hexagonal grid using $2 \times 7$ 3 MHz frequency channels centered at 1365 and 1435 MHz.[4] The cleaning and calibration of the raw data is done using an automated pipeline based largely on routines in the Astronomical Image Processing System (AIPS).[5] It has a typical rms of 0.15 mJy and an angular resolution of 5″ (Becker et al. 1995). The high-frequency NVSS at 1.4 GHz in VLA D configuration covers 82% of the celestial sphere with angular resolution 45″ and a rms of ∼0.45 mJy (Condon et al. 1998). Thus, the FIRST data has a 9× better resolution than the NVSS data. The better resolution and high sensitivity of the FIRST survey (White et al. 1996) allow us the possibility of studying the morphologies of a large number of moderately weak radio galaxies. In our search of HT sources, we use the 2014 December release version of FIRST data, which contains approximately 946,432 radio sources.

### 2.2. Search Strategy

A total of 946,432 radio sources are given in the FIRST catalog. Here, we looked for only HT radio sources with moderate luminosity and size. The better resolution of the FIRST images (∼5″) gives us a sufficient number of sources with a HT structure. First, we separate the sources that have an angular size greater than or equal to 10″ (i.e., at least twice the convolution beam size). To short out the sources, a self-made filtering script was built. A total of 95,243 sources were filtered from the script. Then, we visually looked at every image carefully and separated those candidates whose jets were bent into "C"-, "V"-, or "L"-like structures. We constructed two tables, depending on their degrees of bending, and present them here in this paper. Earlier, Cheung (2007) searched for "X"-shaped radio galaxies from the FIRST survey. He filtered those sources whose sizes are ⩾15″. Bera et al. (2020) also searched for "X"-shaped, or "winged", radio sources from the FIRST survey (2014 data release). They shortlisted the sources with sizes ⩾10″. Here, we have also selected those candidates whose single-component extension ⩾10″. The selected HT sources have a single component, where most of our cases have more than one component.

#### 2.2.1. Separating the Narrow-angle Tail and Wide-angle Tail Sources

Depending on their bending angle between the two radio jets, HT sources are classified into two subclasses, NAT and WAT. Sources which have a bending angle less than 90° are defined as NAT sources and sources which have a bending

---

[4] http://sundog.stsci.edu/first/description.html
[5] http://info.cv.nrao.edu/aips/

**Table 1**
Candidate Wide-angle Tail Radio Sources

| Catalog Number | Name | R.A. (J2000.0) | Decl. (J2000.0) | Ref. | Redshift ($z$) | $F_{1400}$ (mJy) | $F_{150}$ (mJy) | $\alpha_{150}^{1400}$ | $L$ (erg s$^{-1}$) ($\times 10^{43}$) | FR Type (I/II) | Other Catalogs |
|---|---|---|---|---|---|---|---|---|---|---|---|
| 1 | J0001−0826 | 00 01 15.11 | −08 26 46.2 | SDSS | 0.33[a] | 156 | ... | ... | ... | ... | ... |
| 2 | J0007+0536 | 00 07 05.88 | +05 36 02.3 | EE | ... | 307 | ... | ... | ... | I | 8 |
| 3[a] | J0017−0149 | 00 17 32.28 | −01 49 23.7 | EE | ... | 91 | ... | ... | ... | II | ... |
| 4 | J0020−0950 | 00 20 08.01 | −09 50 53.3 | EE | ... | 11 | ... | ... | ... | I | 1 |
| 5 | J0027+0224 | 00 27 19.11 | +02 24 28.9 | 2MASX | 0.13[b] | 133 | ... | ... | ... | ... | 1, 2 |
| 6 | J0034+1140 | 03 34 30.50 | +11 40 40.9 | SDSS | ... | 18 | ... | ... | ... | ... | ... |
| 7 | J0035−0834 | 00 35 03.48 | −08 34 39.2 | 2MASS | ... | 387 | ... | ... | ... | II | 8 |
| 8 | J0041−0651 | 00 41 10.52 | −06 51 05.2 | EE | ... | 48 | ... | ... | ... | ... | 1 |
| 9[a,c] | J0041+0028 | 00 41 52.16 | +00 28 34.9 | SDSS | 0.15[a] | 94 | ... | ... | ... | II | 13, 15 |
| 10 | J0044+1026 | 00 44 58.90 | +10 26 44.2 | EE | ... | 226 | ... | ... | ... | ... | ... |

**Notes.** In column 1 and in the machine-readable version, a, b, c, and d denote the sources present in Blanton (2000), Wing & Blanton (2011), Proctor (2011), and Paterno-Mahler et al. (2017), respectively.
**References.** 1: NRAO VLA Sky Survey (NVSS; Condon et al. 1998); 2: VLA Low-Frequency Sky Survey (VLSS; Cohen et al. 2007); 3: 3C (Bennett 1962; Edge et al. 1959); 4: 4C (Pilkington & Scott 1965; Gower et al. 1967; Caswell & Crowther 1969); 5: 5C (Kenderdine et al. 1966; Pooley & Kenderdine 1968; Pooley 1969; Willson 1970; Pearson 1975; Waggett 1977; Schuch 1981; Benn et al. 1982); 6: 6C (Baldwin et al. 1985; Hales et al. 1988, 1990, 1991, 1993a, 1993b); 7: 7C (McGilchrist et al. 1990; Kollgaard et al. 1994; Waldram et al. 1996; Vessey & Green 1998); 8: Parkes-MIT-NRAO Radio Survey (PMN; Griffith et al. 1994); 9: The Parkes Catalogue of Radio Sources (PKS; Bolton et al. 1964); 10: Texas Survey of Radio Sources (TXS; Douglas et al. 1996); 11: Cul (Slee 1995); 12: 87 GB (Gregory & Condon 1991); 13: Automatic Spectroscopic K-means-based classification (ASK; Sánchez et al. 2011); 14: 2 Micron All Sky Survey Extended objects—Final Release (2MASX; Skrutskie et al. 2006); 15: Galaxy Evolution Explorer All-Sky Survey Source Catalog (GALEXASC; Agüeros et al. 2005); 16: GALaxy Evolution Explorer Medium Imaging Survey Catalog (GALEXMSC; Agüeros et al. 2005); 17: Gaussian Mixture Brightest Cluster Galaxy (GMBCG; Jiangang et al. 2010); 18: Gaussian Mixture Brightest Cluster Galaxy (MaxBCG; Koester et al. 2007); 19: New General Catalogue (NGC; Dreyer 1888); 20: B2 (Colla et al. 1970, 1972, 1973; Fanaroff & Riley 1974); 21: B3 (Ficarra et al. 1985).
[a] Represents spectroscopic redshifts.
[b] Represents photometric redshifts.

(This table is available in its entirety in machine-readable form.)

angle greater and equal to 90° are defined as WAT sources. The angle measurement was done by taking the angle that the two radio jets' outward vector (direction) made at the center (optical counterpart or eye estimated). Due to projection effects or the orientation of the sky, a WAT may appear as a NAT. Here, we use the one and only parameter angle as the defining criterion due to the fact that we can measure the angle for all sources. However, on the other hand, the luminosity is not available for all of the sources due to the unavailability of redshifts.

### 2.2.2. Finding the Optical Counterparts and Properties

The optical counterparts of the HT sources were selected visually, based on the relative position of radio morphology and optical source. The optical counterparts are selected by using SDSS data from its data release 12 (DR12; Alam et al. 2015), which is the final data release of the SDSS-III, the Digital Sky Survey (DSS), and the NASA/IPAC Extra-galactic Database.[6] An active galactic nucleus (AGN) is located at the central core, from which radio jets are emitted. Generally, an optical galaxy is found in the central region and is known as an optical host galaxy. In some cases there are no optical counterparts found, in which case we then use our best-guessed "eye-estimated" position as the central spot for those sources. We overlaid the FIRST image of the source on the respective DSS2 red image in gray scale. Here we used a search radius 2′ to locate the possible optical counterpart. For WATs, optical counterparts were found for 210 sources out of a total of 430 sources. For NATs, optical counterparts were found for 149 sources out of a total of 287 sources.

### 3. Result

We report a total of 717 new HT sources, in which 430 are WAT and 287 are NAT radio sources, from the FIRST survey. With the available radio and optical data, we calculate the spectral indices and radio luminosities of all HT candidates. We also study the statistical properties of these candidates.

All the WAT and NAT sources are listed in Tables 1 and 2, respectively. In Figures 1 and 2 we present sample images of 12 WAT and NAT sources, respectively. In addition, in Figure 3 we draw a histogram presenting the angle distributions of the NAT and WAT sources. All sources are listed in their respective table according to the R.A. (J2000.0). In column 6, we estimate the redshift of the candidates from the SDSS, Two Micron All-Sky Survey eXtended (2MASX), and Two Micron All Sky Survey (2MASS) catalogs. Redshifts were found for 261 out of 717 HT sources; among them, 153 WATs and 108 NATs. In column 7, we list the flux densities of the sources from NVSS data. FIRST and NVSS both observed the sky at 20 cm (1400 MHz). We chose NVSS flux instead of FIRST flux because of the following two reasons. (i) NVSS has a low resolution (with a resolution of 45″ with 1″–7″ astrometric accuracy) VLA D configuration compared to the B-configuration FIRST survey (with a resolution of 5″ with astrometrical errors of 0.″5–1″). Therefore, FIRST distinguishes small-scale structures with accurate positions, but underestimates flux for extended sources. NVSS has less accurate positional measurements and cannot distinguish small-scale structures, but measures more accurate fluxes for extended sources and can find low-surface-brightness objects missed by FIRST. (ii) Due to its high resolution and lack of antennas in short spacing, the FIRST survey is susceptible to

---
[6] https://ned.ipac.caltech.edu

Table 2
Candidate Narrow-angle Tail Radio Sources

| Catalog Number | Name | R.A. (J2000.0) | Decl. (J2000.0) | Ref. | Redshift (z) | $F_{1400}$ (mJy) | $F_{150}$ (mJy) | $\alpha_{150}^{1400}$ | $L$ (erg s$^{-1}$) ($\times 10^{43}$) | FR Type (I/II) | Other Catalogs |
|---|---|---|---|---|---|---|---|---|---|---|---|
| 1 | J0012−0607 | 00 12 48.62 | −06 07 03.1 | 2MASX | 0.20[a] | 66 | ⋯ | ⋯ | ⋯ | II | 15 |
| 2 | J0017+0827 | 00 17 37.70 | +08 27 52.9 | EE | ⋯ | 55 | ⋯ | ⋯ | ⋯ | I | ⋯ |
| 3 | J0020+0004 | 00 20 14.18 | +00 04 48.0 | EE | ⋯ | 170 | 457 | −0.44 | ⋯ | I | 16 |
| 4 | J0023+0717 | 00 23 46.06 | +07 17 55.5 | 2MASX | 0.25[a] | 74 | ⋯ | ⋯ | ⋯ | I | 1, 2, 10 |
| 5[c] | J0041−0925 | 00 41 49.95 | −09 25 48.6 | GALEX | ⋯ | 35 | ⋯ | ⋯ | ⋯ | II | ⋯ |
| 6 | J0047+1034 | 00 47 11.24 | +10 34 58.2 | 2MASS | ⋯ | 24 | ⋯ | ⋯ | ⋯ | ⋯ | ⋯ |
| 7 | J0054+0302 | 00 54 43.83 | +03 02 10.5 | SDSS | 0.41[a] | 18 | 192 | −1.06 | 1.58 | ⋯ | 1 |
| 8 | J0056−0119 | 00 56 30.55 | −01 19 19.9 | EE | ⋯ | 162 | ⋯ | ⋯ | ⋯ | ⋯ | ⋯ |
| 9[a,c] | J0056−0120 | 00 56 02.70 | −01 20 03.5 | CGCG | 0.04[a] | 370 | ⋯ | ⋯ | ⋯ | II | 8, 11, 14, 16 |
| 10 | J0111+1141 | 01 11 40.13 | +11 41 34.1 | EE | ⋯ | 13 | ⋯ | ⋯ | ⋯ | II | ⋯ |

**Notes.** In column 1 and in the machine-readable version, a, b, c, and d denote the sources present in Blanton (2000), Wing & Blanton (2011), Proctor (2011), and Paterno-Mahler et al. (2017), respectively.
**References.** 1: NRAO VLA Sky Survey (NVSS; Condon et al. 1998); 2: VLA Low-Frequency Sky Survey (VLSS; Cohen et al. 2007); 3: 3C (Bennett 1962; Edge et al. 1959); 4: 4C (Pilkington & Scott 1965; Gower et al. 1967; Caswell & Crowther 1969); 5: 5C (Kenderdine et al. 1966; Pooley & Kenderdine 1968; Pooley 1969; Willson 1970; Pearson 1975; Waggett 1977; Schuch 1981; Benn et al. 1982); 6: 6C (Baldwin et al. 1985; Hales et al. 1988, 1990, 1991, 1993a, 1993b); 7: 7C (McGilchrist et al. 1990; Kollgaard et al. 1994; Waldram et al. 1996; Vessey & Green 1998); 8: Parkes-MIT-NRAO Radio Survey (PMN; Griffith et al. 1994); 9: The Parkes Catalogue of Radio Sources (PKS; Bolton et al. 1964); 10: Texas Survey of Radio Sources (TXS; Douglas et al. 1996); 11: Cul (Slee 1995); 12: 87 GB (Gregory & Condon 1991); 13: Automatic Spectroscopic K-means-based classification (ASK; Sánchez et al. 2011); 14: 2 Micron All Sky Survey Extended objects—Final Release (2MASX; Skrutskie et al. 2006); 15: Galaxy Evolution Explorer All-Sky Survey Source Catalog (GALEXASC; Agüeros et al. 2005); 16: Galaxy Evolution Explorer Medium Imaging Survey Catalog (GALEXMSC; Agüeros et al. 2005); 17: Gaussian Mixture Brightest Cluster Galaxy (GMBCG; Jiangang et al. 2010); 18: Gaussian Mixture Brightest Cluster Galaxy (MaxBCG; Koester et al. 2007); 19: New General Catalogue (NGC; Dreyer 1888); 20: B2 (Colla et al. 1970, 1972, 1973; Fanaroff & Riley 1974); 21: B3 (Ficarra et al. 1985).
[a] Represents spectroscopic redshifts.

(This table is available in its entirety in machine-readable form.)

flux-density loss. Two-point spectral indexes ($\alpha_{150}^{1400}$) between 150 MHz (TIFR GMRT Sky Survey) and 1400 MHz (NVSS) are computed in column 9. In column 10, we have also calculated the luminosity of the HT sources with known redshifts. Using our eyes estimation, the morphological types of NATs and WATs are represented in column 11: "I" represents FR-I type and "II" represents FR-II type.

Of the 717 sources in our sample, 77 are in the Blanton (2000) sample (marked as "a"), 42 are in the Wing & Blanton (2011) sample (marked as "b"), 103 (32 WAT and 71 NAT) are in the Proctor (2011) sample (marked as "c"), and 24 are in the Paterno-Mahler et al. (2017) sample (marked as "d"), as noted in Tables 1 and 2.

### 3.1. Spectral Index ($\alpha_{150}^{1400}$)

The two-point spectral indexes ($\alpha_{150}^{1400}$) of the NATs and WATs are calculated using equation $S_\nu \propto \nu^\alpha$, where $S_\nu$ is the radiative flux density at a given frequency, $\nu$, and $\alpha$ is the spectral index. The available spectral indexes for the WATs and NATs are 195 and 112, respectively, and are listed in Tables 1 and 2, respectively. Due to the higher rms in the TGSS images, the rest of the sources were not detectable in the 150 MHz map of TGSS. Figure 4 shows the spectral index ($\alpha_{150}^{1400}$) distribution of 112 NAT sources. The histogram shows that the total span of $\alpha_{150}^{1400}$ is from −1.33 to −0.11 for the NAT candidates. Among the NAT sources, J1235+0741 has the lowest spectral index, with $\alpha_{150}^{1400} = -1.33$, and J0734+3611 has the highest spctral index, with $\alpha_{150}^{1400} = -0.11$. The histogram shows a peak near $\alpha_{150}^{1400} = -0.65$. Similarly, Figure 5 shows the spectral index ($\alpha_{150}^{1400}$) distribution of 195 WAT sources. The histogram shows that the total span of $\alpha_{150}^{1400}$ is from −1.27 to −0.15. Among the WAT sources, J1300+2916 has the lowest spectral index, with $\alpha_{150}^{1400} = -1.27$, and J1457+0232 has the highest spectral index, with $\alpha_{150}^{1400} = -0.15$. The histogram shows two peaks near $\alpha_{150}^{1400} = -0.60$ and −0.75.

The mean and median values of the spectral index ($\alpha_{150}^{1400}$) of NAT sources are −0.62. Similarly, for WAT sources the mean and median values are −0.67. This means the spectral index values of our newly discovered NAT and WAT sources are similar to normal-sized radio galaxies (Oort et al. 1988; Gruppioni et al. 1997; Kapahi et al. 1998; Ishwara-Chandra et al. 2010; Mahony et al. 2016). In the work of Oort et al. (1988), they surveyed the Lynx field with the Westerbork Synthesis Radio Telescope (WSRT) at 325 and 1400 MHz. Gruppioni et al. (1997) surveyed the Marano field using the Australia Telescope Compact Array (ATCA) at 1.4 and 2.4 GHz. Kapahi et al. (1998) surveyed the Molonglo Radio Catalogue sources with the VLA in the $L$ and $S$ bands. In the work of Ishwara-Chandra et al. (2010) they surveyed the LBDS-Lynx field by using the GMRT at 150 MHz and other archival data from the GMRT at 325 MHz and 610 MHz frequency, along with data from other surveys such as WENSS, NVSS, and FIRST. Mahony et al. (2016) surveyed the Lockman Hole field using LOFAR 150 MHz and WSRT 1.4 GHz.

### 3.2. Radio Luminosity ($L_{rad}$)

The radio luminosity ($L_{rad}$) is the characteristic parameter that defines the strength of the radio jets. One can easily estimate the surface brightness of an extragalactic radio source from the radio luminosity by some empirical relations (Heeschen 1966). The FR dichotomy is done depending on the radio luminosity value.

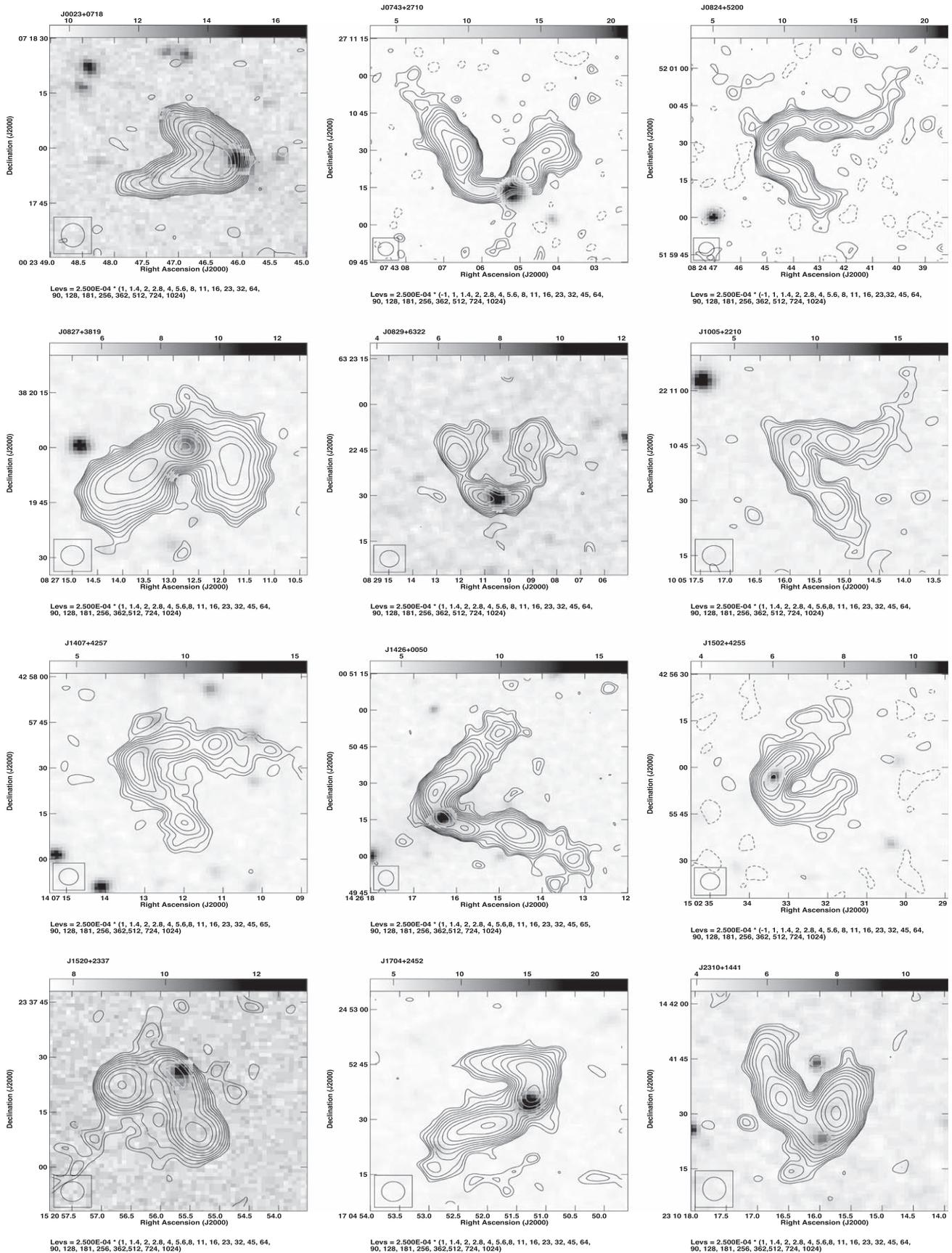

**Figure 1.** FIRST image of a sample of 12 narrow-angle tail (NAT) radio sources (contours) overlaid on the DSS2 red image (gray scale).

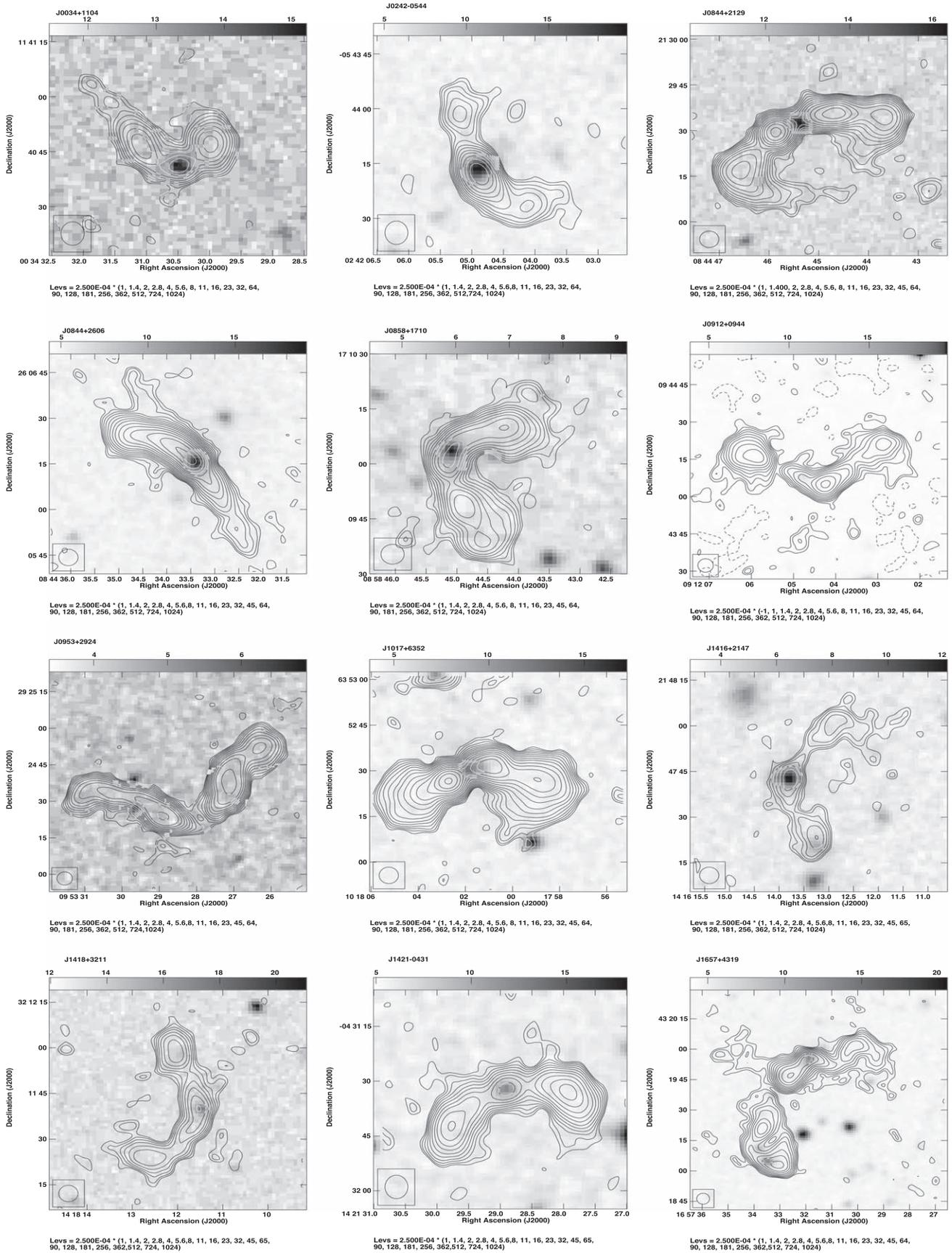

**Figure 2.** FIRST image of a sample of 12 wide-angle tail (WAT) radio sources (contours) overlaid on the DSS2 red image (gray scale).

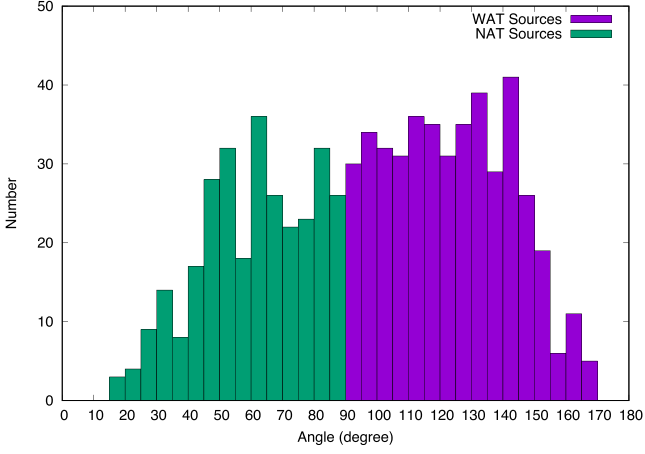

Figure 3. Histogram showing the angle distribution of narrow-angle tail (NAT) and wide-angle tail (WAT) sources. Left panel is for NAT sources (green color) and right panel is for WAT sources (violet color).

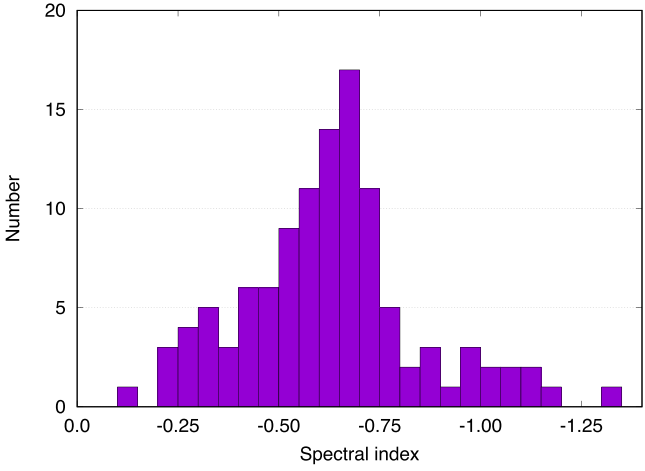

Figure 4. Histogram showing spectral index ($\alpha_{150}^{1400}$) distribution of NAT candidates.

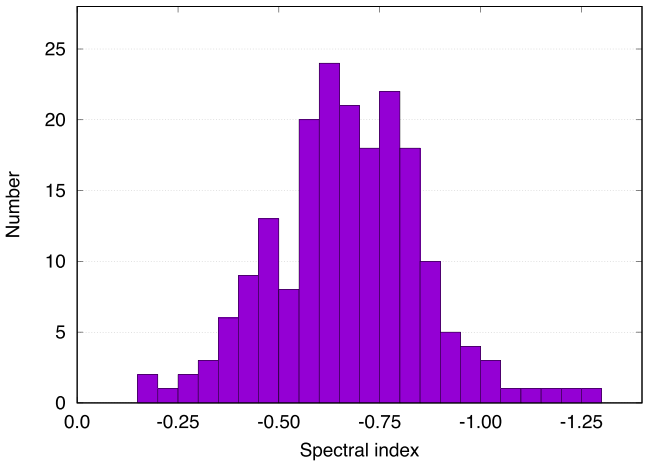

Figure 5. Histogram showing spectral index ($\alpha_{150}^{1400}$) distribution of WAT candidates.

The radio luminosity ($L_{\rm rad}$) is calculated from the equation

$$L_{\rm rad} = 1.2 \times 10^{27} D_{\rm Mpc}^2 S_0 \nu_0^{-\alpha} (1+z)^{-(1+\alpha)} \\ \times (\nu_u^{(1+\alpha)} - \nu_l^{(1+\alpha)})(1+\alpha)^{-1} {\rm erg\ s}^{-1}, \quad (1)$$

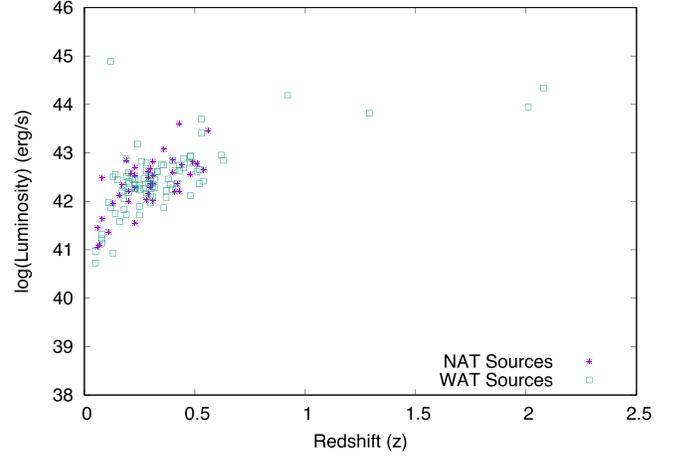

Figure 6. Plot showing the radio luminosity ($L_{\rm rad}$) distribution of WAT and NAT sources with redshifts ($z$).

where $D_{\rm Mpc}$ is the luminosity distance to the source (Mpc), $S_0$ is the flux density (Jy) at a given frequency $\nu_0$ (Hz), $z$ is the redshift of the radio galaxy, $\alpha$ is the spectral index ($S \propto \nu^\alpha$), and $\nu_u$ (Hz) and $\nu_l$ (Hz) are the upper and lower cut-off frequencies (O'Dea & Owen 1987). In our calculation, we assume upper and lower cut-off frequencies of 100 GHz and 10 MHz, respectively.

In Figure 6, we plot the radio luminosity distribution of the NATs and WATs presented in our sample with known redshifts ($z$). Note that in this figure the data points are overcrowded in a particular area of the figure. The apparent increase of luminosity with redshift is largely a result of Malmquist bias. A total 129 data points, of which 42 are NATs and 87 WATs, are plotted here. The radio luminosities of the sources at 1400 MHz frequency is in the order of $10^{43}$ erg s$^{-1}$, which is quite similar to a typical radio galaxy.

The most luminous NAT source is J2348+1157, with $L_{\rm rad} = 3.96 \times 10^{44}$ erg s$^{-1}$, and the most luminous WAT source is J1607+4825, with $L_{\rm rad} = 2.20 \times 10^{45}$ erg s$^{-1}$. Similarly, the least luminous NAT source is J1353+3305, with $L_{\rm rad} = 1.13 \times 10^{42}$ erg s$^{-1}$, and the least luminous WAT source is J1253+0604, with $L_{\rm rad} = 1.47 \times 10^{38}$ erg s$^{-1}$. For WAT sources, the mean log$L$ (erg s$^{-1}$) and median log$L$ (erg s$^{-1}$) values of luminosity are 1.03 and 0.37, respectively. Similarly, for NAT sources, the mean log$L$ (erg s$^{-1}$) and median log$L$ (erg s$^{-1}$) values of luminosity are 0.66 and 0.43, respectively.

### 3.3. Fanaroff–Riley Classes

Here, the classification of the FR types is based on visual inspection. We individually check for the flux-density distribution of each selected HT radio source. We check for whether the source flux density is greater at the central region than at the edges or whether the edges are more radio luminous than the center. For the edge-brightened cases, we marked them as FR-II type and for the edge-darkened and diffuse types, we identified them as FR-I type. If no conclusion is reached on the relative flux in between the edges and the central core, we did not classify them as FR-I or FR-II type. Note that we did not use the luminosity criterion in this classification. Among all the HTs from our catalog, 266 (37%) sources are FR-I type and 317 (44%) are FR-II type.

## 4. Discussion

The redshifts are estimated for 108 out of 287 (38%) NAT sources. Among all the NAT candidates, six sources have redshift values greater than 0.5 ($z \geqslant 5$). J1418+0515 has the highest redshift value, 0.583845 ±0.000178, and J1334+2537 has the lowest redshift value, of 0.04, among all NATs. Similarly, redshifts are known for 153 WATs out of 430 (36%). Fourteen candidates have redshift values greater than 0.5 ($z \geqslant 5$) among all WAT sources. The source J1051+0051 has the highest redshift value, 2.00660 ±0.00500 (Croom et al. 2004), among all WATs as well as all HT candidates in our catalogs. J1250+1133 is another high-redshift WAT source, with redshift $z = 1.28500 \pm 0.00000$ (Richards et al. 2009). J1253+0604 is the closest WAT source, with redshift $z = 0.000255$.

The brightest HT source is J1143+5201, with flux ($F_{1400}$) 2541 mJy. This source is also the brightest source among all WATs. The brightest NAT source is J1217+0340, with flux ($F_{1400}$) 1225 mJy. Higher-resolution observations will help us to study the dynamics as well as different aspects of such kinds of sources.

## 5. Conclusion

We visually examined each of the 95,243 selected sources from the last data release (2014 December 17) of the VLA FIRST survey at 1400 MHz. Based on the morphology, we found a total of 717 HT sources from the sample. A few of them are single component and most of them are multi-component. Depending on the bending angle of the two lobes, we classified them as NAT or WAT. Among all HT sources, 287 were NAT and 480 were WAT, respectively. We visually cross-matched our HT sample with SDSS data for the optical counterpart. Optical counterparts were found for 359 sources, 149 for NATs and 210 for WATs. Based on visual inspection of the flux distribution along with the sources, we classified our HT sources into two categories, FR-I and FR-II. Among all HTs, 266 (∼37%) sources are FR-I type and 317 (∼44%) sources are FR-II type. For 134 (∼19%) sources, we could not classify them as either FR-I or FR-II due to their complex flux distributions. A steep spectral index between 150 and 1400 MHz was found for 95 (∼75%) of the HT sources.

Various authors have previously cataloged BT sources on the basis of pattern-recognition programs by computer or via visual inspection. Several sources which were not included in the previous catalogs are included in this present catalog. Our catalog is the first one that contains the highest number of HT sources (717) to date by only visual search. This catalog of HT sources will help in future different statistical studies.

Owen & Rudnick (1976) first identified bent radio sources, which they found in the Abell cluster. Thereafter, in most identifications, we have found that HT radio sources are located in such cluster environments. Blanton (2000) cataloged 384 bent-double sources by visually examining a sample of 3200 multicomponent sources from the 1997 April data release catalog of the FIRST survey. The author used BT samples as a tracer of the galaxy cluster. Within the complete sample area, they found a total of 102 clusters in the Abell catalog. In their sample, bent-double sources are found in ∼5% of clusters. To examine optical environments around double-lobed radio sources, Wing & Blanton (2011) cross-correlated their radio samples with the SDSS. They showed that among all the samples, visually selected bent radio sources (as compared to straight or single-component sources) are more often associated with galaxy clusters, with 78% of visually selected bent sources found in cluster environments. Paterno-Mahler et al. (2017) searched galaxy clusters from the surrounding fields of 646 bent, double-lobed radio galaxies from the Clusters Occupied by Bent Radio AGN survey. Based on galaxy overdensity measurements, they found 190 galaxy-cluster candidates (most at high redshift) from the surrounding fields of their sample. So, bent radio sources are a good tracer to find rich galaxy clusters. In this context of finding galaxy clusters, our catalog will be very useful.

### ORCID iDs

Tapan K. Sasmal  https://orcid.org/0000-0001-9251-9456
Soumen Bera  https://orcid.org/0000-0001-7121-4258